\documentclass{article}
\usepackage[T1]{fontenc} 
\usepackage[utf8]{inputenc} 
\usepackage{ismir,amsmath,cite,url}
\usepackage{graphicx}
\usepackage{color}
\usepackage[ruled,vlined]{algorithm2e}
\usepackage{booktabs}
\usepackage{multirow}
\usepackage{subfigure}
\usepackage{lineno}

\usepackage{xcolor}
\usepackage{ulem}
\title{Semi-supervised learning using teacher-student models for vocal melody extraction}

\multauthor
{Sangeun Kum$^1$ \hspace{1cm} Jing-Hua Lin$^2$ \hspace{1cm} Li Su$^2$ \hspace{1cm} Juhan Nam$^1$} { \\
$^1$ Graduate School of Culture Technology, KAIST, South Korea \\
$^2$ Institute of Information Science, Academia Sinica, Taiwan\\
{\tt\small keums@kaist.ac.kr, jhlin@iis.sinica.edu.tw, lisu@iis.sinica.edu.tw, juhan.nam@kaist.ac.kr }
}
\def\authorname{First author, Second author, Third author, Fourth author}

\begin{document}

\maketitle
\begin{abstract}
The lack of labeled data is a major obstacle in many music information retrieval tasks such as melody extraction, where labeling is extremely laborious or costly. Semi-supervised learning (SSL) provides a solution to alleviate the issue by leveraging a large amount of unlabeled data. In this paper, we propose an SSL method using teacher-student models for vocal melody extraction. The teacher model is pre-trained with labeled data and guides the student model to make identical predictions given unlabeled input in a self-training setting. We examine three setups of teacher-student models with different data augmentation schemes and loss functions. Also, considering the scarcity of labeled data in the test phase, we artificially generate large-scale testing data with pitch labels from unlabeled data using an analysis-synthesis method. The results show that the SSL method significantly increases the performance against supervised learning only and the improvement depends on the teacher-student models, the size of unlabeled data, the number of self-training iterations, and other training details. We also find that it is essential to ensure that the unlabeled audio has vocal parts. Finally, we show that the proposed SSL method enables a baseline convolutional recurrent neural network model to achieve performance comparable to state-of-the-arts.

\end{abstract}

\section{Introduction}\label{sec:introduction} 

One of the key elements in the success of deep learning is a large amount of labeled data. However, when the labeled data is scarce in a given task, it can be a bottleneck in leveraging the power of deep neural networks. The issue has been found in many music information retrieval (MIR) tasks as well. Among others, melody extraction research has suffered from it as pitch labeling requires experienced annotators to handle the annotation tool and the process is extremely labor-intensive \cite{salamons2019}.

The lack of labeled data in melody extraction research has been tackled in several different ways. A popular method to alleviate the issue is data augmentation which increases labeled data by transforming the input audio, for example, using pitch-shifting~\cite{kum2016melody,lu2018vocal,kum2019joint}. Data augmentation, however, has the limitation in covering the diversity in the input space. Another approach is using multi-track audio data \cite{hsu2009improvement,bittner2014medleydb,chan2015vocal}. This allows to use monophonic pitch tracking algorithms for the melodic source and therefore it expedites laborious the pitch labeling. However, multi-track recording datasets often maintain individual tracks as stem files where multiple similar sound sources can be mixed (e.g., main vocal and backing vocal). Therefore, obtaining clean pitch labels from multi-track audio can be not straightforward~\cite{salamon2017analysis,bittner2017deep}. Recently, melody MIDI files, which are more easily accessible, have been utilized to guide melody extraction from audio with transfer learning techniques from the symbolic to audio domain~\cite{lu2018vocal,gao2019vocal}. MIDI data exhibit greater flexibility than audio on data augmentation, but still face limitations on representing natural pitch contours of singing voice, which usually contain subtle variations such as vibrato and portamento.

Semi-supervised learning (SSL) is another but more general strategy to address the lack of labeled data. SSL uses a large amount of unlabeled data, which is usually easy to collect, jointly with labeled data. A popular class of SSL methods is based on self-training in the teacher-student framework. 
Recent works have combined random data augmentation with the SSL methods to encourage the model to produce robust output even when input is perturbed. This approach has achieved state-of-the-art performance on image classification~\cite{berthelot2019mixmatch,xie2019self,sohn2020fixmatch}, speech recognition~\cite{movsner2019improving}, and audio classification~\cite{lu2019semi}. 
There are a few MIR researches that used the teacher-student framework to address the lack of labeled data, for example, in automatic drum transcription~\cite{wu2017automatic} and singing voice detection ~\cite{schluter2016learning,meseguer2018dali}. 
However, to the best of our knowledge, recent advances in SSL methods that leverage the power of deep neural networks and random data augmentation in the teacher-student framework have been not studied yet in the music domain.  

In this paper, we apply the SSL methods to vocal melody extraction with the following contributions. First, we present the SSL methods for vocal melody extraction leveraging large-scale unlabeled music datasets. This prevents the model from overfitting to small labeled data and improve the performance. Second, we compare three setups of teacher-student models along with various audio data augmentation techniques. We show the model with the consistency regularization is most effective. Third, we investigate effective SSL strategies by exploring joint training, the size of unlabeled data, and the number of self-training iterations. 
Fourth, we show that the proposed teacher-student training method enables a baseline convolutional recurrent neural network model to achieve performance comparable to state-of-the-arts. 
Finally, apart from the SSL method, we propose large-scale testing data artificially generated from unlabeled data using an analysis-synthesis framework, considering the lack of labeled data even at the testing stage. Evaluation on the diverse and sizable test set will reinforce the effectiveness of the proposed method. 
For reproducibility, the source code and pre-trained model used in this paper are available online\footnote{\url{https://github.com/keums/melodyExtraction_SSL}}.

\section{Related work}\label{sec:related_work}

The teacher-student framework has been previously studied in several MIR tasks to address the lack of labeled data. Wu and Lerch applied the approach to automatic drum transcription~\cite{wu2017automatic}. They used multiple teacher models based on non-negative matrix factorization (NMF) trained with different datasets and a student model based on deep neural network trained with labels from the teachers. They showed that the student model outperforms the teacher models. However, it was not a self-training setting where the teacher model is repeatedly replaced with an improved student model. 
Schl{\"u}ter explored the self-training for singing voice detection~\cite{schluter2016learning}. They first trained a convolutional neural network (CNN) on the original labels with low-granularity, then a second network on pseudo-labels with high-granularity from the first network, and a third network on the summarized saliency maps from the second network. They showed this self-improvement worked up to the third network. However, they conducted the self-training on weakly-labeled data in the context of multiple-instance learning and did not used any unlabeled data. Recently, Meseguer-Brocal et al. used the teacher-student paradigm for singing voice detection to create a large-scale time-aligned vocal melody and lyrics dataset~\cite{meseguer2018dali}. They consistently improved the teacher model by increasing the correlation between the prediction of the model and the time-aligned lyrics annotation.

\section{Methods}\label{sec:methods}

\begin{figure*}[t]
 \centerline{
 \includegraphics[width=0.95\textwidth]{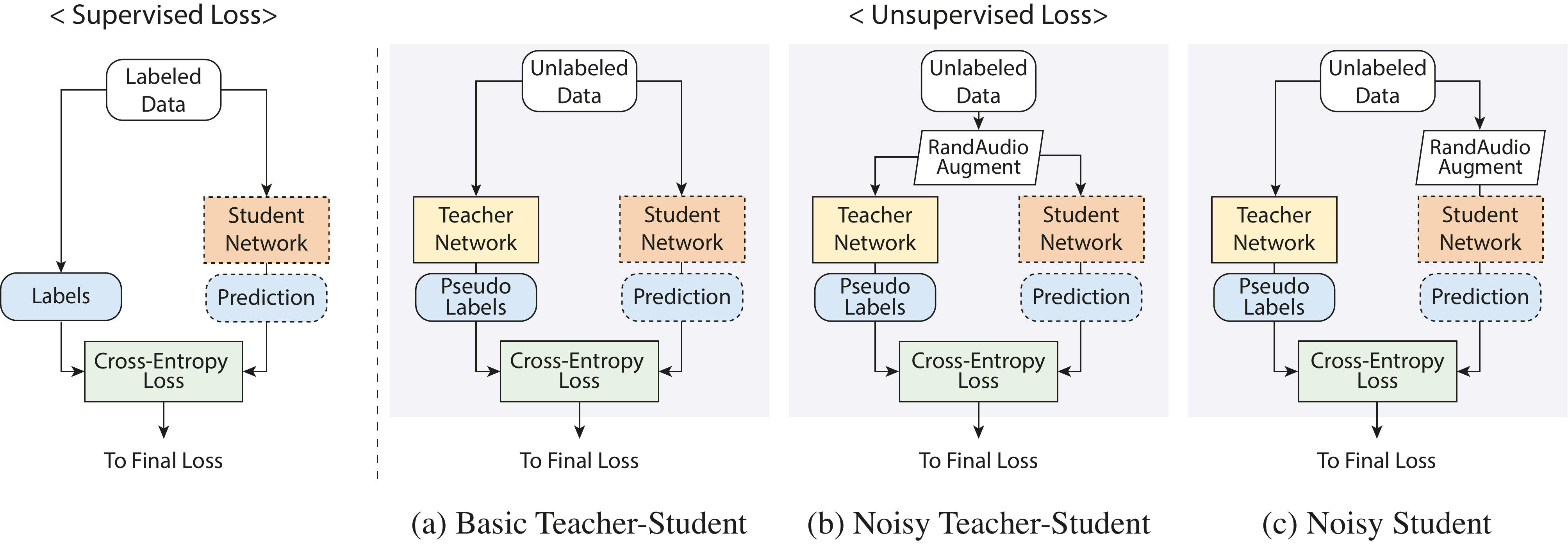}}
 \caption{Diagram of the three Teacher-Student models. }
 \label{fig:TS_model}
\end{figure*}

\subsection{Model Architecture}\label{subsec:architecture}
Recent melody extraction algorithms have used CNN \cite{bittner2017deep,su2018vocal,chen2019cnn}
 and its variants \cite{hsieh2019streamlined,chou2018hybrid,kum2019joint}
as a standard architecture. Since we focus on the effectiveness of SSL in this paper, we employ a previously proposed convolutional recurrent neural network (CRNN) which was a baseline architecture in \cite{kum2019joint}. The CRNN architecture consists of 4 ResNet blocks and a bi-directional long short-term memory layer. 
We first merge the audio waveforms into a mono channel and downsample them to 8 kHz. We then calculate the logarithmic-magnitude spectrogram using short-time Fourier transform with a 1024-point Hann window and an 80-point hop size.
The CRNN architecture takes 31 consecutive frames of the spectrogram as input and predicts a pitch label quantized with a resolution of 1/8 semitone and ranged from E2 (82.4 Hz) to B6 (1975.7 Hz). The size of the output layer is 442, including a non-vocal label.

\subsection{SSL in the Teacher-Student Framework} \label{subsec:SSL_TS}

Our SSL method is based on self-training in the teacher-student framework where the teacher model is first trained with labeled data and then the student model is trained with artificial labels generated from the teacher model using unlabeled data. The artificial labels can be the prediction distribution vector~\cite{berthelot2019mixmatch,xie2019self} or one-hot vector determined by the class with a highest confidence~\cite{lee2013pseudo,sohn2020fixmatch}. We formally describe the overall procedure in Algorithm~\ref{al:algorithms}. We first train the initial teacher model $T_1$ using only labeled data~$\mathcal{D}$ where $x_d$ are labeled examples and $y_d$ are one-hot reference labels. For unlabeled data~$\mathcal{U}$ where $x_u$ are unlabeled examples, we use random data augmentation to generate noisy input data $\mathcal{\tilde{U}}$ where $\tilde{x}_u$ are noisy unlabeled examples. RandAudioAugment (RAA) is an audio version of random data augmentation method which is described in Section~\ref{subsec:data_augmentation}. While it is more effective to use random data augmentation on the student model only in image classification ~\cite{xie2019self}, we also try applying it for both teacher and student models for ablation study. Once we train the student model jointly with the labeled data and unlabeled data (with pseudo labels), we replace the teacher model with the student model. We repeat the same pseudo labeling and the training with a new student model.

\begin{algorithm} [!t]
\SetAlgoLined
 Train a teacher network $T_1$ on labeled data $\mathcal{D} = \{(x_d,y_d): d \in (1,...,N)\}$\;
 Generate augmented data $\mathcal{\tilde{U}} = \{\tilde{x}_u=RAA(x_u): u \in (1,...,M)\} $ from unlabeled data $\mathcal{U} = \{x_u: u \in (1,...,M)\} $\;
 \For {i = 1 to k}{
   Use $T_i$ to generate pseudo labels for $\mathcal{U}$ (or $\mathcal{\tilde{U}}$)\;
   Train student network $S_i$ using both $\mathcal{D}$ and $\mathcal{U}$ (or $\mathcal{\tilde{U}}$) as training data\;
   $T_{i+1}$ = $S_i$\;
 }
    \caption{Train SSL Models}
 \label{al:algorithms}
\end{algorithm}

\subsection{Proposed Teacher-Student Models}\label{subsec:student_models}
Our proposed Teacher-Student models are illustrated in~\figref{fig:TS_model}. 
The supervised loss $\mathcal{L}_D$ is computed with labeled data and defined as:
\begin{equation}\label{eq:loss_d}
\mathcal{L}_D = \frac{1}{N}  \sum_{d=1}^{N} H( y_d, p(y|x_d;\theta_s ))
\end{equation}
where $H(\cdot)$ denotes the cross-entropy between the pitch label $y_d$ and pitch prediction $p(y|x)$, and $\theta_s$ denotes a set of parameters of the student model. The supervised loss is a common loss term of the three investigated teacher-student models. 
Each of them are explained below.  \\ \\
\textbf{Basic Teacher-Student} is a fundamental teacher-student framework that uses the unlabeled data $\mathcal{U}$ but trains the student network with the pseudo labels generated from the teacher network.
The final loss of Basic Teacher-Student $\mathcal{L}_B$ is defined as
\begin{equation}\label{eq:loss_fs}
\mathcal{L}_B = \mathcal{L}_D + \frac{1}{M} \sum_{u=1}^{M} H( y_u, p(y|x_u;\theta_s ))
\end{equation}
where $y_u$ is the pseudo labels on $\mathcal{U}$ generated by the teacher network, i.e. $y_u = p(y|x_u;\theta_t)$ where $\theta_t$ to denote the parameters of teacher network. The basic teacher-student model is illustrated in~\figref{fig:TS_model}(a). \\ \\
\textbf{Noisy Teacher-Student} takes noisy unlabeled data $\mathcal{\tilde{U}}$ for both of the teacher and student networks using RAA and the rest is the same as the basic teacher-student model.  

The final loss of Noisy Teacher-Student $\mathcal{L}_N$ is defined as
\begin{equation}\label{eq:loss_fn}
\mathcal{L}_N = \mathcal{L}_D + \frac{1}{M}  \sum_{u=1}^{M} H( \tilde{y}_u, p(y|\tilde{x}_u;\theta_s ))
\end{equation}
where $\tilde{y}_u$ is a prediction on $\mathcal{\tilde{U}}$ generated by the teacher network, i.e. $\tilde{y}_u = p(y|\tilde{x}_u;\theta_t)$. The noisy teacher-student model is illustrated in~\figref{fig:TS_model}(b). 
\\ \\
\textbf{Noisy Student} takes noisy unlabeled data $\mathcal{\tilde{U}}$ only for the student network while the teacher network takes unnoised input $\mathcal{U}$ to generate the pseudo labels. The idea is that the student should produce consistent outputs that minimize the difference from the teacher even though the input is perturbed~\cite{xie2019self}. This notion is also similar to consistency regularization~\cite{sajjadi2016regularization,xie2019unsupervised}. 

The final loss of Noisy Student $\mathcal{L}_C$ is defined as
\begin{equation}\label{eq:loss_fc}
\mathcal{L}_C = \mathcal{L}_D + \frac{1}{M}  \sum_{u=1}^{M} H( y_u, p(y|\tilde{x}_u;\theta_s ))
\end{equation}
The noisy student model is illustrated in~\figref{fig:TS_model}(c).

\subsection{Data Augmentation}\label{subsec:data_augmentation} 
We conducted pitch-shift by $\pm$ 1,2 semitone on the labeled data~$\mathcal{D}$ (audio and corresponding labels). In the melody extraction task, it has shown that pitch-shifting can improve the generality and performance of the model by increasing the amount of audio and label pairs for different \emph{f$_0$}~\cite{kum2016melody,bittner2018multitask}. For data augmentation of unlabeled data~$\mathcal{U}$, we propose RandAudioAugment (RAA) inspired by RandAugment~\cite{cubuk2019randaugment}, which is a method of randomly applying different kinds of transformations to increase image data. RAA converts audio by randomly selecting multiple audio effects as follows: audio equalizer (low-shelf, high-shelf), filters (low-pass, high-pass), overdrive, phaser, and reverb. 
Here, we use \emph{pysndfx} that is a Python library designed for applying effects to audio files\footnote{https://github.com/carlthome/python-audio-effects}. We sampled a random magnitude of each transformation from a pre-defined range. The implementation details for RAA are also described in the source code.

\subsection{Data Selection}\label{subsec:data_selection}

The SSL algorithm using large-scale unlabeled data may suffer from labeling noise. Unlabeled data are highly likely to have audio without vocals. 
Filtering only high-confidence examples or the top-K examples in image classification has demonstrated to be an effective method to handle the labeling noise~\cite{yalniz2019billion,xie2019self}. Likewise, we performed data selection so that only the tracks with vocal ratios exceeding a threshold were used for training. To estimate the ratio of vocals included in the track, we used our singing voice detector\footnote{https://github.com/keums/SingingVoiceDetection} based on CNN based on~\cite{schluter2015exploring}. 
Considering the distribution of vocal ratio in the FMA, we set the threshold to 0.3.

\vspace{-2mm}
\section{Datasets}\label{sec:dataset} 
\begin{table}
\centering
\resizebox{0.95\columnwidth}{!}{%
\begin{tabular}{@{}cccc@{}}
\toprule
                                                                                & \textbf{Dataset} & \textbf{Number of Tracks} & \textbf{Total Length} \\ \midrule
\multirow{3}{*}{\begin{tabular}[c]{@{}c@{}}Training\\ (Labeled)\end{tabular}}   & RWC              & 100                       & 6h 47m                \\
                      & MedleyDB         & 61               & 2h 39m      \\
                      & iKala       & 262              & 2h 6m       \\ \midrule
\multirow{4}{*}{\begin{tabular}[c]{@{}c@{}}Training\\ (Unlabeled)\end{tabular}} & In-house          & 535                       & 6h 21m                \\
                      & FMA\_small  & 3,521 / 8,000      & 25h / 60h      \\
                      & FMA\_medium & 10,639 / 25,000    & 89h / 208h    \\
                      & FMA\_large  & 40,505 / 106,574 & 337h / 888h \\ \midrule
\multirow{4}{*}{Test} & ADC04       & 12               & 4m          \\
                      & MIREX05     & 9                & 4m          \\
                      & MedleyDB    & 12               & 43m         \\
                      & AST218      & 218              & 14h 53m       \\ \bottomrule
\end{tabular}
}

\caption{Description of datasets. In FMA, 
The two numbers indicate tracks with vocal (the vocal ratio above 0.3) and all tracks respectively.}

\label{tab:dataset}
\end{table}


\tabref{tab:dataset} shows the simple statistics of the labeled and unlabeled training datasets and test datasets. 

\subsection{Labeled Data}\label{subsec:labeled_data} 
We used the three labeled datasets (RWC~\cite{goto2002rwc}, MedleyDB~\cite{bittner2014medleydb}, and iKala~\cite{chan2015vocal}) and split them into a train and validation set following~\cite{bittner2017deep}. We augmented the training data by pitch-shifting with $\pm$ 1,2 semitone. The total length of the labeled training data amounts to about 55 hours after the data augmentation.

\subsection{Unlabeled Data}\label{subsec:unlabeled_data} 
As to unlabeled data, we used an in-house dataset crawled from YouTube and the Free Music Archive (FMA)~\cite{fma_dataset}. 
The in-house dataset is pop songs with vocals recorded in a variety of environments. It includes both public-released and user-uploaded tracks.
FMA is a large-scale open dataset containing up to 106,574 tracks and covers 161 genres of music. We used FMA for performance comparison on data scalability. 
The FMA has three different subsets depending on the number of the track and genre included: FMA\_small (FMA$_S$), FMA\_medium (FMA$_M$), and FMA\_large (FMA$_L$). We selected vocal tracks from them as described in Section~\ref{subsec:data_selection} and denote the selected versions as FMA$_{Sv}$, FMA$_{Mv}$, and FMA$_{Lv}$, respectively. We augmented the unlabeled datasets via RAA during training as described in Section~\ref{subsec:data_augmentation}.

\subsection{Test Data}\label{subsec:test_data} 
\subsubsection{Public Test Sets}\label{subsubsec:public_testset} 
We used three public test sets (ADC04\footnote{http://labrosa.ee.columbia.edu/projects/melody/}, MIREX05, and MedleyDB) to evaluate the performance of vocal melody extraction.
In this study, we excluded non-vocal tracks from ADC04 and MIREX05, and used songs not included in training data for MedleyDB.
To obtain the ground truth for singing voice in MedleyDB, we adopted its 'MELODY2' annotations. 
These three datasets have been commonly used to compare the performance of melody extraction. 
However, the number of tracks and the total length are very limited as shown in \tabref{tab:dataset}. 

\subsubsection{Proposed Large-Scale Test Set}\label{subsubsec:ast218} 
To make up the scarcity of testing data for evaluating singing voice extraction algorithms, we propose a new test set composed of DSD100~\cite{SiSEC16} and MusDB18~\cite{musdb18}. The two multitrack datasets were originally designed for source separation. Each track has four isolated stems: vocals, drums, bass, and others. 
Following the analysis/synthesis framework~\cite{salamon2017analysis}, the singing melodies for 218 selected tracks\footnote{Songs that appear in MedleyDB were excluded for they were part of the training data, but songs in MusDB18 having counterparts in DSD100 were not removed for they are not exactly identical. Additionally, 12 songs that do not have discernible vocal melodies were also excluded.} were synthesized with automatically generated \emph{f$_{0}$} contours. 
In detail, for each song, we extracted the melody of the vocals with five different pitch trackers, and each \emph{f$_{0}$} information along with the vocal audio 
was fed into the WORLD~\cite{morise2016world} (D4C edition~\cite{morise2016d4c}) vocoder to reproduce five monophonic variations of the vocal stem.
The original vocal audio was parameterized into harmonic and aperiodic spectral envelopes, and then resynthesized with provided pitch contours. Then a mask was applied to 
filter intervals without \emph{f$_{0}$} information. For remixing, the amplitude of the synthesized vocal was weighted to that of the original vocal stem, and the rest stems were directly summed up as accompaniments, then mixed with the weighted synthesized vocal that perfectly matched the \emph{f$_{0}$} annotation. These 1,090 polyphonic mixtures with accurate and automatic annotations constitute the proposed analysis/synthesis test set, AST218\footnote{https://sites.google.com/view/mctl/resource}.

Each track in AST218 has five variations whose vocal melody was annotated separately with five different pitch estimators: CREPE~\cite{kim2018crepe} (with confidence threshold of 0.5 and 0.7), pYIN~\cite{mauch2014pyin}, and 
Lu\&Su~\cite{lu2018vocal} (with time step of 10 and 20ms), as they have different merits. 
Since there is no exact way to pinpoint a common optimal confidence threshold across the entire dataset, we chose two different threshold values for CREPE: one is 0.5, suffering from high false positive (FP) but preserving details; the other threshold is 0.7, acceptable FP though sacrificing some recall. 
pYIN was chosen for it has even lower FP while producing stable and continuous melodic lines when the vocal stem is monophonic. 
However, it is not stable in the pholyphonic scenario, which is universal in DSD100 and MusDB18.
In need of other polyphonic-based melody estimators to balance the \emph{f$_{0}$} quality, we chose two time step setups of 
the Lu\&Su model: 20ms, at which this model is optimized; and 10ms, which provides more continuous predictions and offers alternative pitch contours when encountering multiple 

The analysis/synthesis framework has been practiced successfully in evaluating monotonic pitch trackers~\cite{kim2018crepe}. As a sanity check, we evaluated several patchCNN~\cite{su2018vocal} setups on the original and resynthesized ADC04, MIREX05, and MedleyDB. The differences of OA are within $\pm$ 2--5\%, which is acceptable, meaning this framework is also applicable for polyphonic test set generation.

When evaluating vocal extraction algorithms on AST218, we averaged the scores from the five variations. Our pilot study shows that these five pitch contours reach consensus over a majority of frames, while the estimations differ for tricky frames. Rather than manually check on the estimated \emph{f$_{0}$}, we used AST218 in an ensemble manner, fully leveraging the spirit of automatic pitch annotation.

\section{Experiments}\label{sec:experiments}

\subsection{Experimental Setup}\label{subsec:experiments_setpup}
\subsubsection{Training Details}\label{subsec:training_details}
We used the CRNN architecture with residual connections and bi-directional long short-term memory in all experiments. The implementation of the model was consistent with that of the main network of~\cite{kum2019joint}. We trained our models using Adam optimizer for 70 epochs on 2 GPUs. The initial learning rate was set to 0.003 in all the experiments. We used a learning rate schedule that reduces the learning rate by 0.7 times if validation accuracy did not increase within three epochs. The model and the training procedures were implemented using Keras
\footnote{We used Keras 2.3.0, Accessed: 15 May 2020}~\cite{chollet2015keras}. 

\subsubsection{Evaluation}\label{subsec:evaluation}
To evaluate the performance of melody extraction, we mainly used overall accuracy (OA) which combines the accuracy of pitch estimation with voice detection. We also used three metrics raw pitch accuracy (RPA) for pitch estimation, and voicing recall (VR) and voicing false alarm (VFA) for voice detection~\cite{salamon2014melody}. These metric are computed by \emph{mir\_eval}~\cite{raffel2014mir_eval} library designed.

\subsection{Experiment 1: Teacher-Student Models}\label{subsec:study_SM}
\begin{figure}[t]
 \centerline{
 \includegraphics[width=\columnwidth]{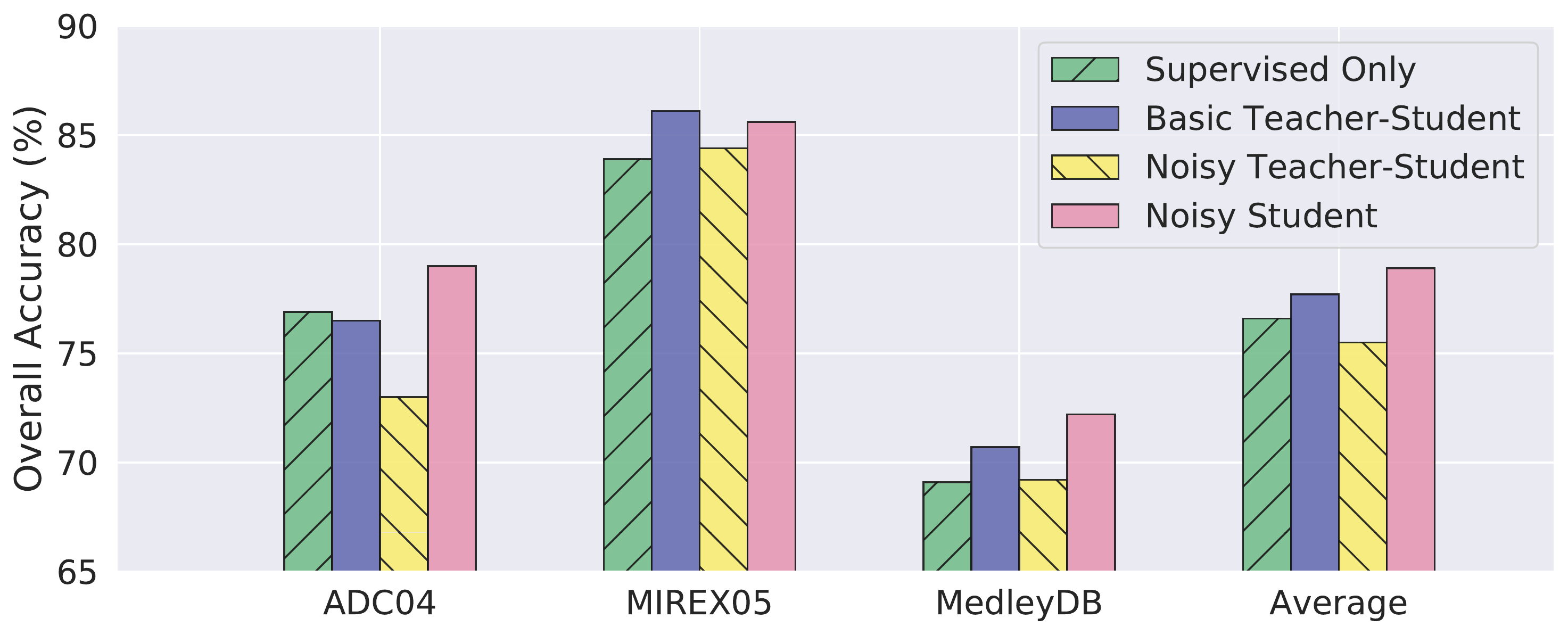}}
 \caption{Comparison with supervised-learning model and three student models on three test sets.}
 \label{fig:result_student}
\end{figure}
Our first experiment is to demonstrate the efficacy of the proposed Teacher-Student models for SSL. 
In this experiment, we trained three Teacher-Student models described in Section~\ref{subsec:student_models} using an in-house dataset as unlabeled data. 
We evaluated the performance of each model on ADC04, MIERX05, and MedleyDB, which have been used as standard test sets for evaluation. As shown in~\figref{fig:result_student}, the basic teacher-student model can achieve 1.1\% higher average OA than the supervised-only model which has 77.7\% average OA. This confirms the possibility of using unlabeled data to improve the performance of melody extraction.
Our experiment also shows that the noisy student model outperforms all the others, having 78.9\% average OA. 

The noisy student model increases OA by 3.1\% with respect to the supervised-only model in MedleyDB, which is especially a challenging dataset because it contains tracks that are difficult to distinguish between vocals and background music, or tracks with excessive audio effects. The results indicate that the student network can be trained reliably using the noisy student model, even if the initial teacher network is not robust to diverse noise. Meanwhile, the performance of the noisy teacher-student has deteriorated, being worse than the supervised-only model. This degradation is probably because the noised teacher model is not generating reliable pseudo labels.  

\subsection{Experiment 2: Joint Training vs. Fine-Tuning}\label{subsec:Study_JT}

\begin{figure}[t]
 \centerline{
 \includegraphics[width=\columnwidth]{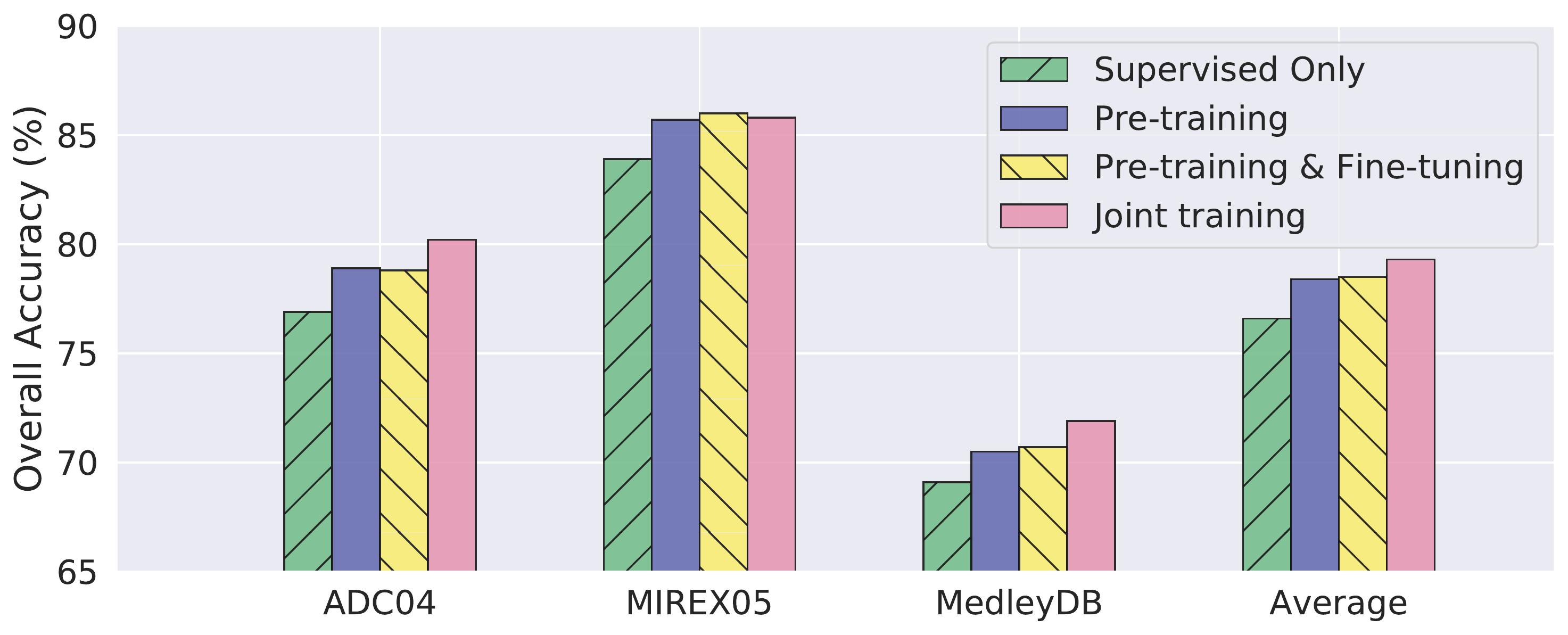}}
 \caption{Comparison with pre-training, fine-tuning, and joint training methods on three test sets.}
 \label{fig:result_joint}
\end{figure}

The training methods of the teacher-student framework can be divided into three approaches depending on how $\mathcal{D}$ and $\mathcal{U}$ are used for training: pre-training on only $\mathcal{U}$ and then fine-tuning on $\mathcal{D}$; joint-training on both $\mathcal{U}$ and $\mathcal{D}$ simultaneously. \figref{fig:result_joint} compares the results among pre-training, fine-tuning, and joint training for the noisy student model.
The jointly trained model achieves 0.8\% higher average OA than the fine-tuned model, with the highest results on MedleyDB.
This indicates that joint training on unlabeled data and labeled data would help the networks produce a decision boundary that better reflects real music~\cite{oliver2018realistic}. Interestingly, the average OA of the pre-trained model only on unlabeled data is higher than that of the supervised learning model. This suggests that the distribution of unlabeled data is similar to that of labeled data. Considering that the in-house dataset consists of pop songs with vocals, the in-house dataset can be seen as having a similar tendency to the labeled data. It provides insight into the data selection in the next experiment.

\subsection{Experiment 3: Size of Training Data}\label{subsec:Study_TD}

\begin{figure}[t]
 \centerline{
 \includegraphics[width=\columnwidth]{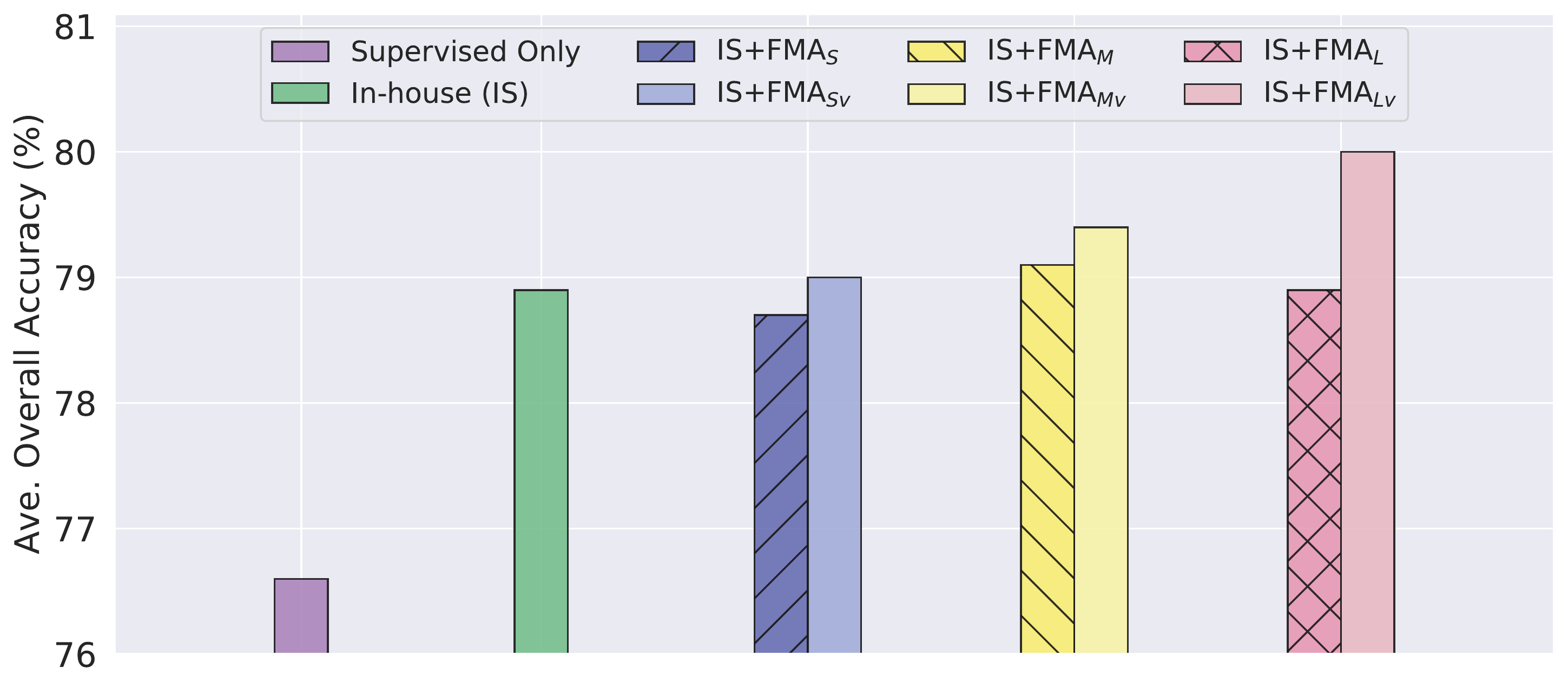}}
 \caption{Comparison with Noisy Students on varied sizes of unlabeled datasets. 
 The subscript `$v$' denotes a selected subset of FMA whose vocal ratio exceeds a threshold. 
 We use the average of OA for three public test sets.
}
 \label{fig:result_dataset}
\end{figure}

We investigated the importance of the size and validity of unlabeled data. To explore the effect of the size of unlabeled data, we started with the in-house dataset as training data for the noisy student model and progressively included larger subsets of FMA. 
The results can be seen in \figref{fig:result_dataset}. 
Although the FMA data set contains more numerous tracks than the in-house dataset, the average OA of $FMA_S$ and $FMA_L$ is lower than that of the model trained only with the in-house dataset.
Note that the proposed model focuses only on vocal melodies. As a result, teacher models may suffer from labeling noise generated by numerous instrument tracks included in the FMA. In addition, all labels on the instrumental track are classified as non-vocal pitch, resulting in data imbalance.

To confirm the validity of the dataset, we performed data selection for each FMA subset as mentioned in Section~\ref{subsec:data_selection} and used them to train each student model.
Interestingly, as the size of the $\mathcal{U}$ increases, the performance of each model tends to be significantly improved. For example, $FMA_{Lv}$ achieves an average OA of 80.2\%, which is 3.6\% higher than the supervised-only model. This indicates that effective SSL requires a large amount of $\mathcal{U}$ with a similar distribution for $\mathcal{D}$.

\subsection{Experiment 4: Iterative Training}\label{subsec:study_IT}

\begin{figure}[t]
 \centerline{
 \includegraphics[width=\columnwidth]{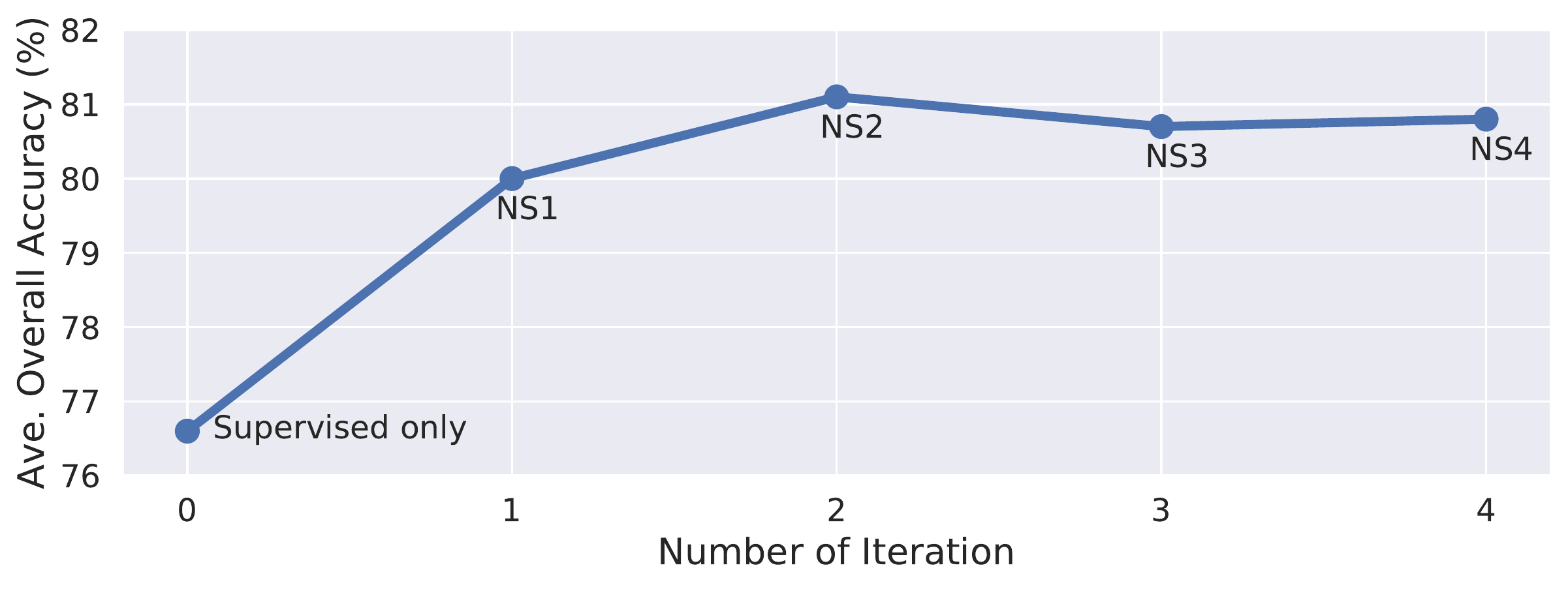}}
 \caption{Effect of iteration training for Noisy Students. 
}
 \label{fig:result_iter}
\end{figure}

We iterated the self-training 4 times for the noisy student model using the in-house dataset and FMA$_{Lv}$. 
The results are illustrated in~\figref{fig:result_iter}. We observe that the performance continuously increases up to 2 iterations achieving the highest average OA of 81.1\%. 
Generally, self-training tends to amplify the error caused by labelling noise during training. However, the noisy student model trained on large-scale unlabeled data can help overcome this difficulty. Nevertheless, increasing the number of training iterations three or more times does not improve performance, and rather slightly lower the accuracy.
 
\subsection{Comparison with State-of-the-Arts}\label{subsec:comparison}


\begin{table}[t]
\resizebox{\columnwidth}{!}{%
\begin{tabular}{@{}ccccc@{}}
\toprule
\textbf{Methods}&               \textbf{ADC04} & \textbf{MIREX05} & \textbf{MedleyDB} & \textbf{AST218} \\ \midrule
    PatchCNN \cite{su2018vocal}   & 76.9 / 72.9          & 69.7 / 73.8            & 44.0 / 59.3             & 42.3 / 59.7           \\
    DSM  \cite{bittner2017deep}   & 89.2 / 72.2          & 87.7 / 80.1            & 80.6 / 75.4             & 38.9 / 68.3           \\
    SegNet \cite{hsieh2019streamlined}  & 88.7 / 83.3    & 82.6 / 80.0            & 70.6 / 75.5             & 41.5 / 68.1           \\
    JDC \cite{kum2019joint}       & \textbf{90.6 / 83.5} & \textbf{91.4 / 87.4}   & 72.7 / 78.1             & 55.8 / \textbf{75.4}  \\\midrule
    Baseline                      & 78.7 / 76.8          & 79.9 / 81.5            & 57.2 / 70.7             & \textbf{56.3} / 69.7           \\
    Proposed (NS)                 & 90.4 / 82.2          & 90.4 / 85.9            & \textbf{76.3 / 79.2}    & 54.2 / 74.2           \\ \bottomrule
\end{tabular}
}
\caption{
Vocal melody extraction results in terms of (RPA / OA) of the proposed and other methods on various test sets. 
The proposed model is iterated the self-training two times using the in-house dataset and FMA$_{Lv}$.
}
\label{tab:comparison}
\vspace{3.7mm}

\resizebox{\columnwidth}{!}{%
\begin{tabular}{@{}ccccc@{}}
\toprule
\textbf{Methods}&               \textbf{ADC04} & \textbf{MIREX05} & \textbf{MedleyDB} & \textbf{AST218} \\ \midrule
    PatchCNN    & 91.8 / 46.1          & 80.3 / \textbf{11.6}            & 60.1 / 22.4             & 61.6 / 26.0           \\
    DSM     & 95.7 / 61.1          & 93.9 / 29.4            & \textbf{85.4} / 26.6             & 44.6 / 7.7           \\
    SegNet  & 95.2 / 38.5    & 92.2 / 24.0            & 78.8 / 21.7             & 51.7 / 10.0           \\
    JDC & 96.7 / 40.2 & \textbf{97.5} / 18.5   & 80.5 / 18.3             & 64.7 / \textbf{8.6}  \\\midrule
    Baseline                      & 92.6 / \textbf{33.8}          & 89.1 / 15.2            & 71.0 / \textbf{16.7}             & \textbf{72.0} / 19.2           \\
    Proposed (NS)                 & \textbf{97.4} / 42.1          & 97.3 / 20.4            & 83.3 / 19.1    & 61.6 / 9.4           \\ \bottomrule
\end{tabular}
}
\caption{Voicing detection results in terms of (VR / VFA) of the proposed and other methods on various test sets. 
}
\label{tab:vcomparison}
\end{table}

We compared the supervised-only model (as a baseline) and proposed the noisy student model (NS) with four recent melody extraction algorithms based on deep neural networks: the patch-based CNN (patchCNN)~\cite{su2018vocal}, the deep salience map (DSM)~\cite{bittner2017deep}, the streamlined encoder/decoder network (segNet)~\cite{hsieh2019streamlined}, and the joint detection and classification model (JDC)~\cite{kum2019joint}, which have open-sourced codes with vocal mode. 
Each method was run with its default parameters, and then evaluated on the three conventional test sets and the newly introduced AST218. Besides, we report the frame-level scores instead of song-level ones to settle uneven song lengths.

\tabref{tab:comparison} and \tabref{tab:vcomparison} list the results of each method on the four test sets. In general, performances of the proposed NS model are comparable to other supervised-learning-based methods and even outperforms others in MedleyDB, and it effectively improves the OA of the baseline by 4.5--8.5\%. The overall rankings of VR and VFA vary across the test sets, but the behavior converges in terms of OA.
One can also observe that the AST218 is the most challenging in the majority of cases. In such a dataset, the performance of the NS model shows that the proposed method is robust to large-scale evaluation.
However, the NS model improves the baseline except for VR and RPA in AST218. This result might be because the simple rule-based remixing of vocal and accompaniment tracks in AST218 is different from the artistic practice of mixing engineers, which can affect voicing detection and, in turn, RPA.       

\section{Conclusion}\label{sec:conclusion}
This study provides a framework of semi-supervised learning using the teacher-student model for vocal melody extraction. 
We compared three setups of teacher-student models and revealed that the NS model is the most effective and robust to real-world music where various noises can be present.
We showed that large-scale unlabeled data is effective when they are properly selected. 
We found that iterative training for the teacher-student model helps improve performance. 
We also confirmed the effectiveness of the proposed method by evaluating it on artificial large-scale test data generated from automatically annotated multitrack data. Although these findings are based only on vocal melody extraction, we believe our method can be extended to other MIR tasks that suffer from the lack of labeled data such as automatic music transcription and chord recognition.

\section{Acknowledgement}
This research was supported by BK21 Plus Postgraduate Organization for Content Science (or BK21 Plus Program) and Basic Science Research Program through the National Research Foundation of Korea (2015R1C1A1A02036962).


\end{document}